# RHIC Power Supply Ramp Diagnostics*

J. T. Morris, T.S. Clifford, B. Frak, J. Laster, A. Marusic, J. van Zeijts
BNL, Upton, NY 11973

## Abstract

Reliable and reproducible performance of the more than 800 Relativistic Heavy Ion Collider (RHIC) magnet power supplies is essential to successful RHIC operation. In order to support power supply commissioning, a system was developed to capture detailed power supply measurements from all the RHIC ring power supplies during acceleration ramps. Diagnostic tools were developed to allow experts to assess ramp reproducibility and rapidly identify problems. The system has now become a routine part of RHIC operations, with data captured for every acceleration ramp. This paper describes the RHIC power supply ramp diagnostic system and considers its impact on RHIC operations.

## 1 INTRODUCTION

More than 800 power supplies are required to operate the superconducting magnets in the Relativistic Heavy Ion Collider (RHIC). Successful operation of RHIC depends on reliable and reproducible performance of these power supplies during acceleration ramps. A system was developed to capture detailed power supply measurements from RHIC ring power supplies during ramps. Diagnostic tools were developed to support expert analysis of this ramp data.

The ramp data sets have proven to be a useful source of information for operators and beam commissioners as well. They provide a quantitative measure of the reproducibility of power supply performance. They are used to analyze power supply response when new modes of ramping are introduced. The ramp data sets are used to identify power supply problems that may go undetected by ordinary power supply alarm mechanisms. Problems in the power supply control system may also be detected by analysis of these data sets.

The ramp diagnostic system has now become a routine part of RHIC operations. Data is captured during every ramp and routinely analyzed by control room personnel. Diagnostic tools have been expanded to simplify data analysis and identify problems quickly. A watching mechanism has also been introduced to bring problems to the immediate attention of operators.

## 2 DATA ACQUISITION

RHIC power supply references are controlled with Wave Form Generators (WFGs), which deliver new values to all of the RHIC power supplies at a 720Hz rate. Multiplexed Analog to Digital Converters (MADCs) [1] provide measurements of a number of power supply parameters including currents and references. These MADCs are configured to capture data at the same 720Hz rate used by the WFGs.

Once a second, the 720Hz WFG and MADC data is read by RHIC Controls Front End Computers (FECs) [2]. The reading of the data is triggered by the 1Hz event on the RHIC Event Link (REL) [3] so that it is precisely synchronized for the WFGs and MADCs for all power supplies. The data is made available for live monitoring programs and stored in 10 second history buffers that are used for post mortem analysis after power supply quenches. The data is also averaged and added to extended history buffers that are used for ramp diagnostics. In order to accommodate a full acceleration ramp, the extended history buffers are currently configured to hold 260 seconds of 30Hz averaged data.

At the end of a RHIC acceleration ramp, the Sequencer program [4][5] triggers a RHIC Event Link event to start the process of saving ramp data in files. When power supply FECs see this event, they simultaneously halt the accumulation of data in their extended history buffers. A console level server process gathers the contents of all of these frozen 30Hz history buffers from the FECs and stores the data in a set of SDDS files [6]. When acquisition is complete, the server process triggers another event which restarts the accumulation of history data.

The term *snapshot* has been used to describe the capture of similar data for individual power supplies. The complete set of power supply data for a RHIC acceleration ramp is called a *snapramp*. The *snapramp* data set includes all

{*} Work supported by US Department of Energy

of the available signals for each power supply. In order to facilitate rapid analysis of *snapramp* data, a compact version of the *snapramp* data set is also created. Three compact ramp data set files are created by 1Hz sampling of the WFG setpoints, the MADC measured current, and the MADC measured reference from the 30Hz *snapramp* data sets.

# 3 POSTRAMP ANALYSIS

## 3.1 Viewing data by supply

*Snapramp* data can be viewed with the *PMViewer* tool, which is also used to view RHIC power supply post mortem data [7]. *Barshow* is another program that has been specifically designed to do graphic comparisons of power supply waveforms for ramp analysis. The user of the *barshow* program selects individual power supplies for analysis. The *barshow* display includes all available signals from the power supply. In addition, calculated difference signals are added to the display.

## 3.2 Comparing ramp data

Users of *barshow* or *PMViewer* must decide which power supplies should be examined. The *pscompare* program is used to do a comparison of ramp data sets for all power supplies, rapidly identifying the power supplies that require further examination. The *pscompare* user selects two ramp data sets. WFG setpoints are typically compared with MADC measured currents or references from the same ramp. Corresponding data sets from different ramps may also be compared. For example, measured currents from a ramp may be compared with measured currents from the preceding ramp.

The results of the comparison are presented in a sorted table as shown in Figure 1. Differences are typically listed as a percentage of the full range of operation for that power supply. The power supply with the largest difference is listed at the top of the table. Filtering options allow a user to limit the presentation to a selected set of power supplies (e.g. all sextupoles or all correctors in alcove 9). For performance reasons, *pscompare* comparisons are typically performed on the compact 1Hz *snapramp* data sets. This has proven to be adequate to identify almost all power supply or control problems. An optional comparison of the full 30Hz *snapramp* data is available but it is significantly slower.

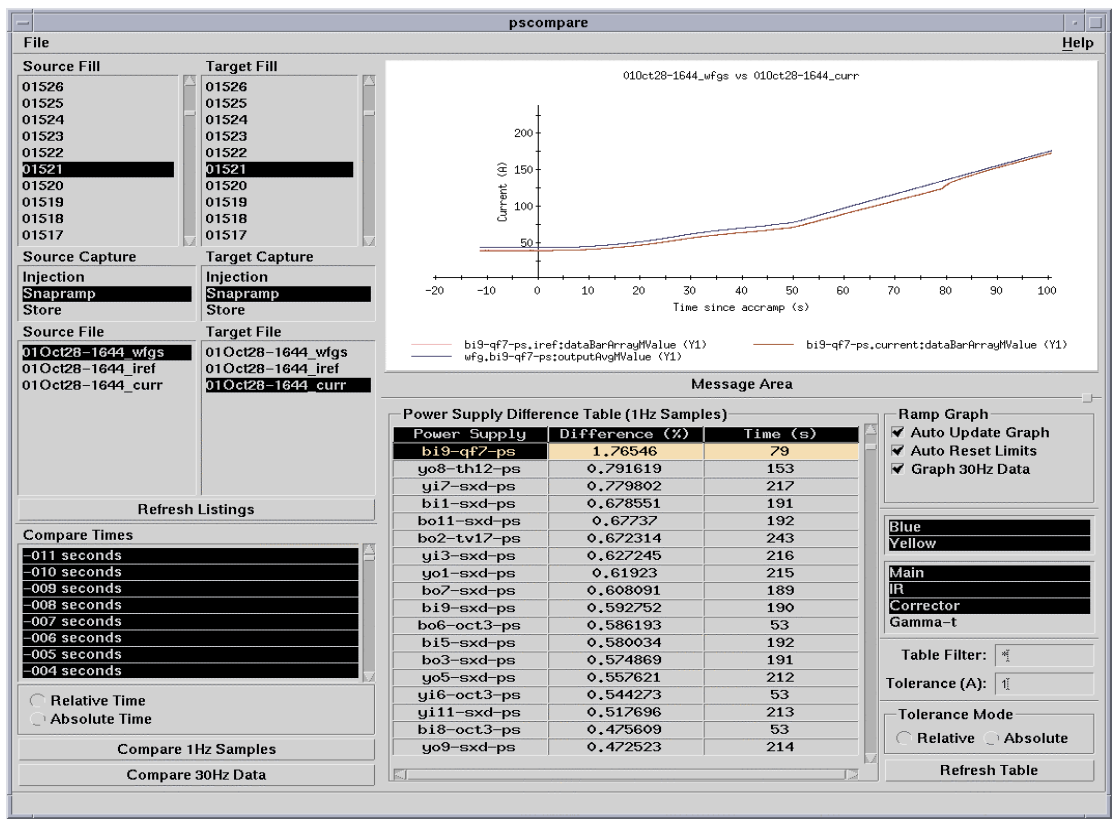

Figure 1: The *pscompare* application with problem power supply highlighted and graphed.

As a user clicks on power supply names in the sorted table, data for that power supply is presented in the graphics area above the table as shown in Figure 3. The default graphic view is a quick plot of 1Hz data for the ramp data sets being compared. A user may optionally display the full 30Hz *snapramp* data for the selected supply. This display may optionally include all available signals from the power supply.

## 4 ALARMS

As described above in the Data Acquisition section, RHIC power supply FECs read and process WFG and MADC data once each second. The data available to the FEC for each power supply includes the current state of the power supply, the WFG setting, an MADC measure of the power supply reference, and an MADC measure of the actual current in the power supply. Software in the FEC compares these values for each power supply that is in an ON state. If these values differ by more than a specified tolerance, an alarm is reported to the RHIC alarm screen in the Main Control Room. During RHIC acceleration ramps, these alarms may be fleeting as some power supplies may only temporarily fall outside tolerance limits. At the end of each ramp, therefore, the RHIC Sequencer displays a record of all alarm conditions that occurred during the ramp.

## 5 FUTURE

Power supply ramp diagnostics have been an essential tool during RHIC commissioning and they are expected to continue to be important for future RHIC operations. The emphasis to date has been on identification of problems by comparing measured currents with expected values. Under more stable operating conditions, comparison of power supply currents with currents measured during a designated reference ramp will be expected.

The data acquisition mechanism is now mature and has been performing very reliably. The system requires significant storage space with 120MB of data in each *snapramp* data set. The demands of this system and the RHIC post mortem system are forcing a significant expansion of data storage capabilities. Only modest changes are expected in data presentation tools.

The alarm capabilities require the most attention. Alarming has been useful but has not realized its full potential. Tolerances have been loosely set to accommodate calibration offsets in MADC measured values. False alarms still are generated due to intermittent measurement errors and a failure to properly detect state when power supplies trip off. A concerted effort needs to be made to correct calibration errors, eliminate all other sources of false alarms, and then set tolerances to the appropriate levels for each power supply.